# Capturing Evidence From Wireless Internet Services Development

Fabio Bella, Jürgen Münch, Alexis Ocampo

**Abstract**—The merging of the Internet with the Wireless services domain has created a potential market whose characteristics are new technologies and time-to-market pressure. The lack of knowledge about new technologies and the need to be competitive in a short time demand that software organizations learn quickly about this domain and its characteristics. Additionally, the effects of development techniques in this context need to be understood. Learning from previous experiences in such a changing environment demands a clear understanding of the evidence to be captured, and how it could be used in the future. This article presents definitions of quantitative and qualitative evidence, and templates for capturing such evidence in a systematic way. Such templates were used in the context of two pilot projects dealing with the development of Wireless Internet Services.

**Index Terms**— Wireless Internet Services, Experience, Evidence, Process Models, Product Models, Baselines, Lessons Learned.

——————————— ◆ ———————————

## 1 INTRODUCTION

The Wireless Internet Services domain is an upcoming application domain that can be characterized as follows: Quickly evolving technology, upcoming new devices, new communication protocols, support for new different media types, varying and limited communication bandwidth, together with the need for new business models that will fit in with the completely new services portfolio. Examples of new wireless Internet services can be expected in the domains of mobile entertainment, telemedicine, travel services, tracking and monitoring services, or mobile trading services.

Due to its recentness, this domain lacks explicit evidence related to technologies, techniques, and suitable business models that could be drivers for software organizations that are willing to be part of the market with new services.

Time to market is important for being competitive, but so is the quality of produced services. Learning from previous experience can help organizations to accomplish these goals.

One problem lies in the reticence of software organizations to capture this knowledge due to the effort and discipline required. Organizations do not want to follow very strict processes or collect information, because time pressure does not allow this.

In this article, we demonstrate that through constant interaction with the developers of the service on defining the goals for capturing evidence, defining the procedures to collect it, and defining not so stringent means for capturing this evidence, valuable context-sensitive evidence can be collected even in a constantly changing domain.

This article presents how evidence gathered from the development of pilot services was explicity described. The means to collect such evidence and the plans to reuse it are discussed, too. The pilot services were developed within the scope of the WISE project.

## 2 THE WISE PROJECT

The work to be described was conducted in the context of the WISE project (Wireless Internet Software Engineering), which started in 2001 and will run through 2004. The project aims at delivering methodologies and technologies to develop services on the wireless Internet.

The WISE project follows an underlying experimental paradigm: Experimenting with methodology and technology in real life applications is seen as the key to understanding, validating, and improving methodology and technology.

In the WISE project, pilots are a means for designing processes and understanding the technology and methodology to engineer and operate with Wireless Internet Services in realistic contexts and different application domains. Based on market demands (such as the need to adapt existing services for the Internet to Wireless Internet Services or to create new services) and company interests, two target contexts were defined: the development of a Wireless Internet service for mobile online trading and the development of a service for Mobile Entertainment.

The industrial partners responsible for the pilot development and the underlying infrastructure are Motorola Global Software Group - Italy (Motorola GSG-Italy), VTT Electronics (Finland), and Investnet (Italy). The industrial partners identified the following success factors for wireless Internet services: time-to-market and the ability to quickly deliver functionality with simultaneous fulfillment of high quality requirements and high usability requirements in terms of service performance.

## 3 CAPTURED EVIDENCE

We applied different experience models to the WISE pilot projects (see Fig. 1), which are an adaption of basic principles of the Experience Factory [1] and QIP [2] approaches.

All kinds of software engineering experience are regarded as experience elements, especially models (such as process models, product models, quality models), instances (such as process

———————————————
*F. Bella, J. Münch, and A.Ocampo are with the Fraunhofer Institute for Experimental Software Engineering, 67661 Kaiserslautern, Germany. Email: {bella, muench, ocampo@ iese.fhg.de).*

*Manuscript received (January 16, 2004).*

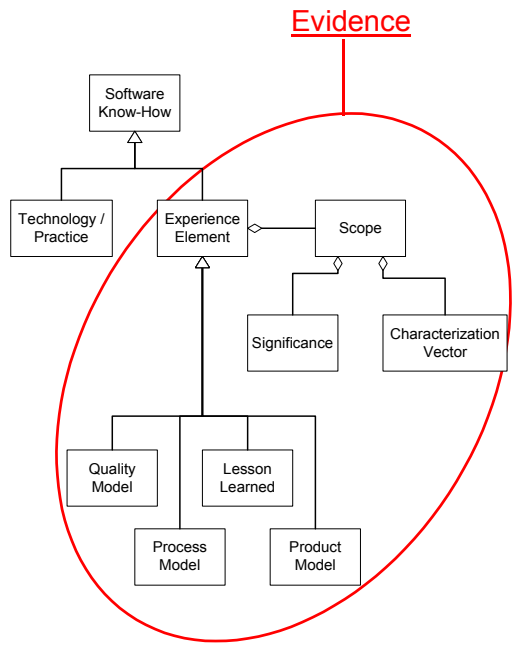

**Fig. 1.** UML diagram for software know-how [6]

traces, products, measurement data, techniques, tools), and qualitative experience (such as lessons learned). For each experience element, the scope of its validity is described. The scope consists of a characterization vector and the significance.

The characterization vector characterizes the environment in which the experience element is valid (see Table 1). The significance describes how the experience element has been validated and to which extent (e.g., validation through formal experiments, single case study, or survey).

**Table 1**
**Excerpt of a Characterization Vector**

| Customization factor | Characteristic | Pilot X |
|---|---|---|
| Domain characteristics | Application type | Computation-intensive system |
| | Business area | Mobile online entertainment services |
| Development characteristics | Project type | Client New development Server New development |
| | Transport protocol | GSM/GPRS/UMTS |
| | Implementation language | Client: J2ME Server: J2EE |
| | Role | Technology provider, service developer |

The proposed modeling allows, for example, to formulate the following facts:

- There is an evidence with the significance *s* that the technology *t* was applied within the context *c* with the result *r*, where *s* is an instance of the class Significance, *t* is an instance of the class Technology, *c* is an instance of the class Characterization Vector, and *r* is an instance of the class Quality Model.
- There is an evidence with the significance *s* that the process model *p* was followed within the context *c*, where *s* is an instance of the class Significance, *p* is an instance of the class Process Model, and *c* is an instance of the class Characterization Vector.
- There is an evidence with the significance *s* that the problem *p* arose and was solved within the context *c*, where *s* is an instance of the class Significance, *p* is an instance of the class Lessons Learned, and *c* is an instance of the class Characterization Vector.

In the following sections, the different models proposed are discussed in more detail.

## 4 PROCESS AND PRODUCT MODELS

Software of good quality can be achieved through the use of systematic development processes and engineering principles. Such issues are also important for producing good quality wireless Internet services.

The question is: How do we produce a software development process suitable for wireless Internet services? Some of the possibilities include: Looking at similar domains and trying to adapt their processes, eliciting the actual practices and tailoring them to the new domain, or the combination of both. In the context of the WISE project, a method was developed and used in order to design an adaptable software process based on existing practices from related domains, industrial piloting, and expert knowledge [3].

As part of the method, descriptive process modeling was used in order to capture the real experience of the process performers and to verify the evidence (products). Process models for each pilot were built through interviews with the process performers.

The instrumentation of the process was done with the metrics of the accompanying measurement programs. The goal of the measurement programs was to baseline key figures as effort distribution.

Data collection sheets were consistent with the activities of the process models, and suitable for the roles described by process participants. More details of the measurement program will be given in the next section. The evidence captured through descriptive process modeling was made explicit with the SPEARMINT™ notation [4].

Experience from related domains was captured through a literature search focused on techniques, tools, methods, case studies, and real projects.

In order to package this experience and make it reusable, the descriptive models for the pilots, and the practices and processes from related fields are integrated into a comprehensive process model. It is planned that this comprehensive process model will be used as the basis for a discussion with the pilot developers in order to create a Reference Model for the domain.

Fig. 2. shows an excerpt of the comprehensive process model that was created.

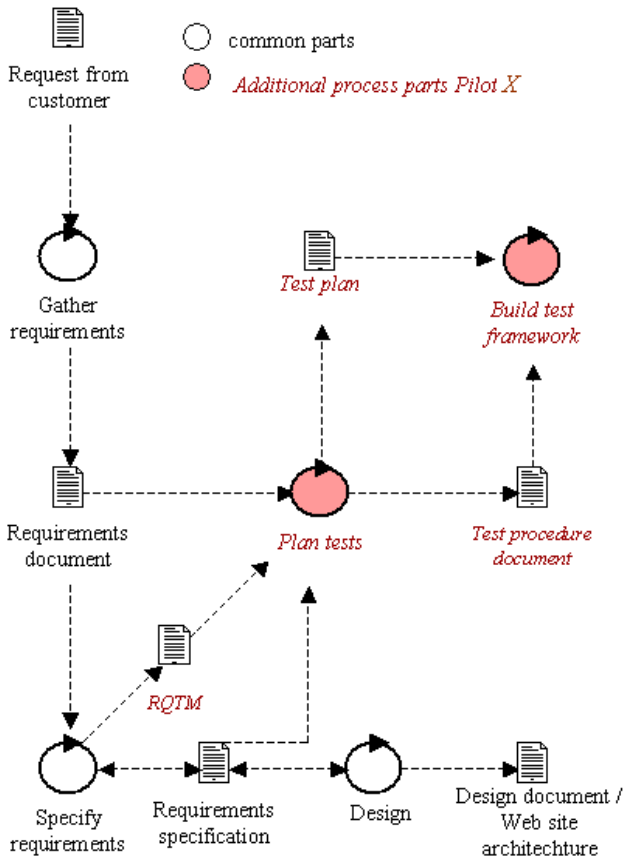

**Fig. 2. Excerpt of comprehensive process and product model**

## 5 QUALITY MODELS

According to [1], effective measurement requires that it (1) specifies the goals by itself, (2) traces these goals to the data, and (3) provides a framework for interpreting these goals in order to understand them.

For the WISE pilots, we used the Goal/Question/Metric (GQM) approach for defining and evaluating a set of operational goals. Briand et al. [5] describe this approach in terms of six major steps:

- Step 1: Characterize the pilot environment.
- Step 2: Identify measurement goals and develop measurement plans.
- Step 3: Define data collection procedures.
- Step 4: Collect, analyze and interpret data.
- Step 5: Perform post-mortem analysis and interpret data.
- Step 6: Package experience.

During the first two steps, business and improvement goals were analyzed and metrics defined according to the process model describing the whole development process as is. The results of this first phase were GQM plans that comprised all metrics defined.

In the following step, the project plan and the process model were used to determine by whom, when, and how data were to be collected according to the metrics. The data collection procedures were the results of this instrumentation.

Raw data were collected according to the data collection procedures. The collected raw data were analyzed and interpreted according to the GQM plan and the feedback provided by the interested parties.

In the fifth step, the interested parties drew consequences according to the analysis and their interpretations.

Finally, analysis, interpretations, and consequences were resumed in the measurement results and collected as experience in the experience database.

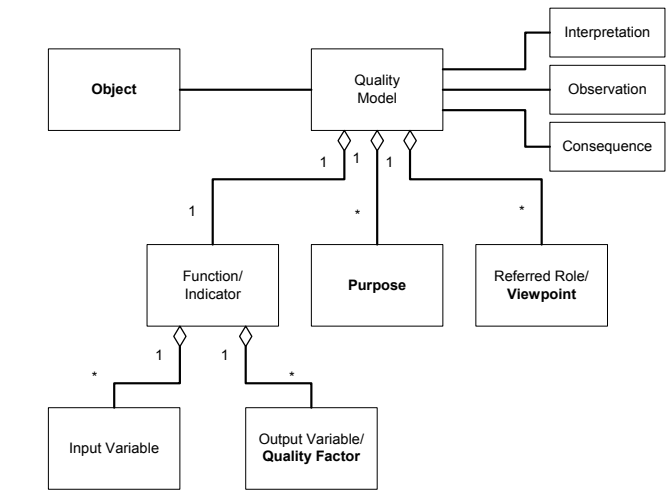

**Fig. 3. UML Diagram for Quality Model [6]**

Fig. 3. highlights the relationship occurring between a quality model and the main aspects of a GQM goal (i.e., object, purpose, quality focus, viewpoint, and context): a quality model refers to the analysis of an object (e.g., software development process) for a certain purpose (e.g., characterization) with respect to a quality focus from a viewpoint (e.g., manager) in a certain context (e.g., WISE service pilot X, first iteration). Furthermore, the figure shows relationships between a quality model and observations, interpretations, and consequences.

Three different types of quality models can be distinguished: project-oriented (e.g., the COCOMO model), process-oriented (e.g., effort distribution model), and product-oriented (e.g., defect slippage model).

According to Fig.3, a quality model aggregates one indicator, i.e., one function tracing the relationship between input variables (e.g., phase identifiers) and output variables (e.g., effort spent on each phase). Indicators are suitable for answering the questions of a GQM plan. Once raw data have been collected and validated, indicators are computed.

Data interpretation rests upon the indicators and the analysis previously performed. It is important to notice, though, that selecting the right indicators in order to know what evidence to collect is a difficult task.

Indicators are not to be seen as the final results of a measurement program, but rather as intermediate results that provide an objective basis for further interpretation.

The following example explains the relationship between measurement goals, questions, measures, and indicators.

- Measurement goal: Effort characterization
- Question: What is the effort distribution broken down by phases?
- Measures:
  o Phase identifier

- o Effort in hours
- Indicator: Effort distribution (broken down by phases).

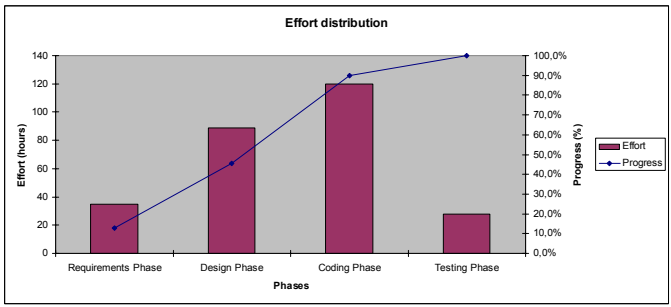

**Fig. 4. Indicator for effort distribution**

In order to package the quality models obtained, a template was used in the context of the WISE project (see Table 2).

**TABLE 2**
**Example of the quality model template**

| Model Id. | WISE-QM3PXI1 |
|---|---|
| Model Name | Effort Characterization Pilot X Iteration 1 Server Side |
| Model Type | Quality model/process-oriented/effort model |
| Significance | 1 Case study |
| Measurement Period | 7-22-2001 – 12-31-2002 |
| Object | Software development process |
| Purpose | Characterization |
| Viewpoint | Manager |
| Characterization Vector/Context | WISE pilot service X, 1st iteration – for further details, see characterization vector CV1PXI1 |
| Indicator | What is the effort distribution (broken down by phases)? 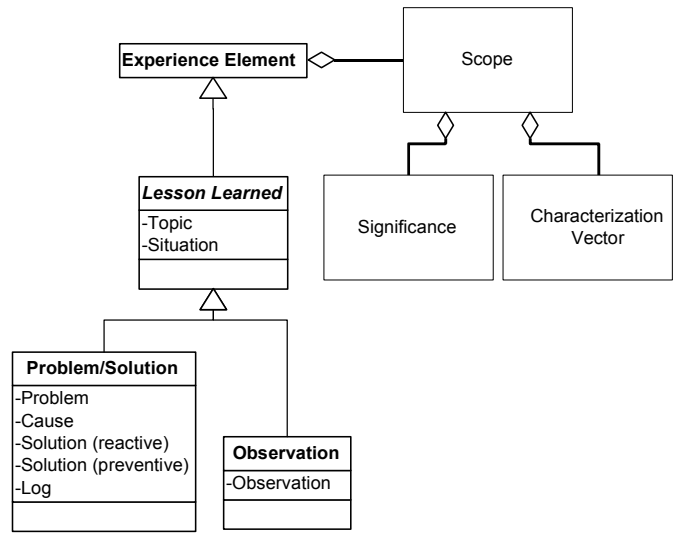 |
| Observations | O1: Lowest effort is spent on requirements phase. O2: More effort than planned is spent on the coding phase. |
| Interpretations | I1 (O1): An external requirements specification was used. I2 (O2): Experience in the platform used is very low: Developers were basically JAVA programmers (1 year experience). I3 (O2): Client-Server interaction was not properly defined. A lot of effort was spent on defining it. First attempts based on TCP/IP did not lead to the performance needed. Therefore, a new transport protocol was defined on the basis of UDP. |
| Consequences | C1 (I3): Try to provide a solution based on TCP/IP at least for UMTS clients. |
| References | Process Model Pilot X Iteration 1 (PM1PXI1) |
| Additional Documentation | D8-V1 "Evaluation - Indicators" |

## 6 LESSONS LEARNED

There are interesting questions to be addressed during a meeting in order to collect lessons learned. Norman L. Kerth has found the following four questions to be keys to focusing an organization on learning and improvement [7]:

- What did we do well, which we might forget if we don't discuss it?
- What did we learn?
- What should we do differently next time?
- What still puzzles us?

**Fig. 5. UML Diagram for Lessons Learned**

In the WISE project, we collected lessons learned at the end of the pilot projects by interviewing project participants. The following definition was used for explicity documenting lessons learned (see Fig.5.):

- **Topic:** This attribute collects relevant keywords. The keywords should be chosen with a keyword-oriented search of the lesson learned in mind.
- **Situation:** This is a brief description of the lesson and of the context in which it was learned.
- **Significance:** Since a project can be seen as a trial within the scope of a wider experiment, significance means the frequency of occurrence of the prob-lem/observation and the control provided over the independent variables (i.e., case study, formal experiment, survey; generally, a single project with a unique characterization vector can be seen as a case study).
- **Characterization Vector / Context:** This attribute collects relevant items from the characterization vector belonging to the project and a reference to the same vector.
- **References:** This is a list of references to other kinds of software know-how (i.e., lessons learned or quality models).
- **Additional Documentation:** This is a list of documents that are relevant for a deeper understanding of the lesson learned or of its consequences. In the case of the WISE project, these are mostly WISE deliverables.

Two different types of lessons learned can be distinguished: A problem / solution pair describes a problem encountered during a project and what was undertaken in order to solve it.
Observations describe experiences that should not be forgotten, since they could be of help for future projects.
The following attributes apply only to problem / solution pairs:
- **Problem:** This is a description of the problem.
- **Cause:** This is a description of events or factors generating the problem.
- **Solution (reactive):** The solution described by this attribute was applied to solve the existing problem.
- **Solution (preventive):** The solution described by this attribute was applied to avoid the problem.
- **Log:** This attribute contains a detailed description of the single steps involved in the solution.

The following attribute applies only to observations:
- **Observation:** This attribute is a textual description of the fact under consideration.

The following is an example of an observation:

TABLE 3
Excerpt of an observation

| Topic | J2ME, WAP 1.0, Push technology, Information system, Cellular phone Nokia 6110 |
|---|---|
| Situation | Advantages of porting an information system from WAP1.0 to J2ME in the case of a cellular phone. |
| Significance | 1 case study |
| Characterization Vector / Context | WISE pilot service X, $2^{nd}$ iteration – for further details, see characterization vector CV3PXI2 |
| Observation | The adoption of J2ME, as an alternative to WAP 1.0 in the case of an information system developed for a cellular phone, can be useful only if push technology is needed to distribute data, and if this technology is supported by the provider. Low level programming of the graphic interface is required to obtain major enhancements compared to WAP1.0, but this increases the load on the terminal. |

The following is an example of a problem/solution pair:

TABLE 4
Excerpt of a problem/solution pair

| Topic | TCP/IP, UDP, transport protocol, real time application, GPRS, J2ME |
|---|---|
| Situation | Phase: Requirement Phase - Pilot service X, iteration 1, developed a multi-client game on GPRS devices supporting J2ME. Since the game is a real time application, very fast communication between client and server is required. First attempts showed that TCP/IP is not fast enough to support the communication between client and server. |
| Significance | 1 case study |
| Characterization Vector / Context | WISE pilot service X, $1^{st}$ iteration – For further details see characterization vector and influence factors as described in D9-V1. |
| Problem | Communication between client (GPRS device) and server side too slow for real time application (game with several clients) |
| Cause | Communication between client and server through TCP/IP transport protocol too slow |
| Solution (reactive) | See preventive solution. |
| Solution (preventive) | A new transport protocol based on UDP was implemented. This solution restricts portability, since many devices supporting J2ME do not support UDP. |
| References | Process Model Pilot X Iteration 1 (PM1PXI1) |
| Additional Documentation | D2-V1 "Methodology - Service Engineering Process". |

## 7 SUMMARY AND OUTLOOK

Developing software nowadays is a task characterized by time-to-market pressure and high quality expectations. The context that surrounds software projects is constantly changing and evolving due to the fast pace of technology. Software development organizations are facing the challenge to find mechanisms for learning from previous experiences and reusing this information in a systematic way.
The presented work has demonstrated that even in rapidly changing contexts, fast and reliable collection of data is possible.
In this work, evidence was defined as a set of process models, product models, quality models, and lessons learned, within a given context.
Gaining evidence in the context of the WISE pilot project showed that it was difficult to formulate appropriate characterization vectors because many of the factors that may influence the projects were only assumed to be relevant by subjective judgement. On the other hand, deviations in the context vectors gave valuable hints on how to tailor these models to the specific project characteristics.
Descriptive process modeling was used successfully as a means for capturing process and product models.
The exact definition of measurement goals and activities was possible through interviews and constant interaction with project participants, and allowed precise acquisition of the relevant measurement data.

The capturing of lessons learned was also successful because of the constant involvment of pilot participants. The templates that were used as a means for packaging lessons learned and quality models were also designed with the feedback from pilot participants. This was helpful during the analysis and interpretation sessions.

The packaging of process models revealed a lack of appropriate mechanisms for coping with variants of experience elements (e.g., notations are missing for describing variants in process models).

Non-anticipated events (such as replanning activities) should be carefully documented, e.g., with a process trace, so that they can be considered during analysis. An analysis of the threats to validity has to be performed in order to become aware of the unknown factors that may influence the results without our knowledge.

Models should be extended with (empirically derived) guidelines on how to adapt the models to project-specific goals and characteristics. Appropriate representation styles for generic knowledge and suitable tailoring mechanisms are needed. A longterm activity should be to investigate the quantitative impact of known factors of the characterization vectors, and to identify new factors.

## ACKNOWLEDGMENT

This work has been funded by the European Commission in the context of the WISE project (No. IST-2000-30028). We would like to thank the WISE consortium, especially the pilot partners, for their fruitful cooperation. We would also like to thank Sonnhild Namingha from the Fraunhofer Institute for Experimental Software Engineering (IESE) for reviewing the first version of the article.